\documentclass[english,5p]{elsarticle}
\usepackage[applemac]{inputenc}
\usepackage{babel}
\usepackage{amsmath}
\usepackage{graphicx}
\usepackage{subfigure}
\usepackage[draft]{hyperref}
\usepackage{color}

\makeatletter

\biboptions{authoryear}

\makeatother

\begin{document}
\begin{frontmatter}

\title{Reply to Chen et al.: Parametric methods for cluster inference perform worse for two-sided t-tests}

\author[mymainaddress,mysecondaryaddress,mythirdaddress]{Anders Eklund\corref{mycorrespondingauthor}}
\cortext[Anders Eklund]{Corresponding author}
\ead{anders.eklund@liu.se}

\author[mymainaddress,mythirdaddress]{Hans Knutsson}

\author[myfourthaddress,myfifthaddress,mysixthaddress]{Thomas E. Nichols}

\address[mymainaddress]{Division of Medical Informatics, Department of Biomedical Engineering,
Link\"{o}ping University, Link\"{o}ping, Sweden}

\address[mysecondaryaddress]{Division of Statistics \& Machine Learning, Department of Computer
and Information Science, Link\"{o}ping University, Link\"{o}ping, Sweden}

\address[mythirdaddress]{Center for Medical Image Science and Visualization (CMIV), Link\"{o}ping University, Link\"{o}ping, Sweden}

\address[myfourthaddress]{Big Data Institute, University of Oxford, Oxford, United Kingdom}

\address[myfifthaddress]{Wellcome Trust Centre for Integrative Neuroimaging (WIN-FMRIB), University of Oxford, Oxford, United Kingdom}

\address[mysixthaddress]{Department of Statistics, University of Warwick, Coventry, United Kingdom}

\begin{abstract}

One-sided t-tests are commonly used in the neuroimaging field, but two-sided tests should be the default unless a researcher has a strong reason for using a one-sided test. Here we extend our previous work on cluster false positive rates, which used one-sided tests, to two-sided tests. Briefly, we found that parametric methods perform worse for two-sided t-tests, and that non-parametric methods perform equally well for one-sided and two-sided tests.

\end{abstract}
\end{frontmatter}

\section{Introduction}

\citet{twosided} discuss an important topic which is often neglected in the neuroimaging field, the use of one-sided or two-sided tests and the lack of multiple comparison correction for two one-sided tests. As mentioned in their paper, in our work on massive empirical evaluation of task fMRI inference methods with resting state fMRI~\citep{eklundPNAS} we used one-sided tests (familywise error rate $\alpha_{\mathrm{FWE}}=0.05$).  We made this choice for two reasons. The first reason was simply that for analyses of randomly created groups of healthy controls, it should make no difference if one uses a one-sided or a two-sided test. The second reason was more practical. FSL and SPM both run one-sided tests by default, and we wished to reflect the typical (if ill-advised) practices of the community. Furthermore, to perform a two-sided permutation test~\citep{winkler}, it would be necessary to run two permutation tests per group analysis (which would double the processing time), since normally only the maximum test value over the brain (or the largest cluster) is saved for every permutation (to form the maximum null distribution).

\section{Methods}

To investigate if performing a two-sided test (as implemented by two tests at $\alpha_{\mathrm{FWE}}  = 0.025$) lead to different false positive rates compared to a single one-sided test (at $\alpha_{\mathrm{FWE}}$ = 0.05), we performed new group analyses for a subset of all the parameter settings used in our previous work~\citep{eklundPNAS,eklundHBM}. Specifically, we only performed two-sample t-tests for the Beijing data~\citep{biswal2}, using 40 subjects (i.e. 20 subjects per group) and a cluster defining threshold of p = 0.001. All group analyses were performed for 4 mm, 6 mm, 8 mm and 10 mm FWHM of smoothing. See our recent work~\citep{eklundHBM} for a description of the six designs (B1, B2, E1, E2, E3, E4) applied to every subject in the first level analysis.

For FSL, group analyses were only performed using FSL OLS, and not using FLAME1 (which is the default option); FLAME1 leads to conservative results if resting state fMRI data is used, while null task fMRI analyses (control-control) with FLAME1 gives FWE rates comparable to FSL OLS~\citep{eklundPNAS}. For AFNI, we used the new ACF (autocorrelation function) option in 3dClustSim~\citep{coxbiorxiv}, which uses a long-tail spatial ACF instead of a Gaussian one. It should be noted that AFNI provides another function for cluster thresholding, ETAC (equitable thresholding and clustering)~\citep{etac}, which may perform better than the long-tail ACF function used here, but we used the ACF approach to be able to compare the two-sided results to our recent work~\citep{eklundHBM}. Contrary to~\citet{twosided}, we did not change the cluster defining threshold to p = 0.0005 when performing two one-sided tests (for SPM, FSL or AFNI), as this represents yet another change in the inference configuration that we rather leave fixed to facilitate the comparison of these results to previous one-sided findings.

\section{Results}

\begin{figure*}
\subfigure[]{
\includegraphics[scale=0.425]{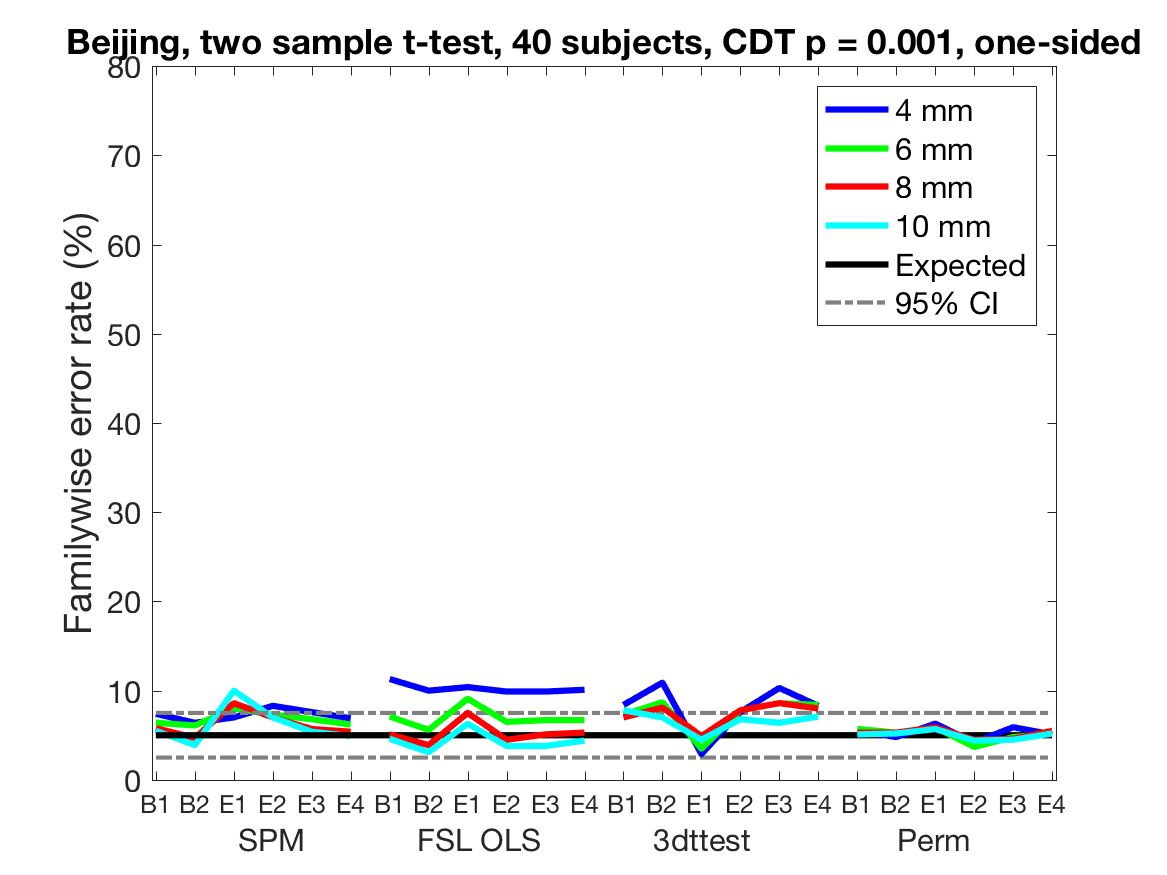}
}
\subfigure[]{
\includegraphics[scale=0.425]{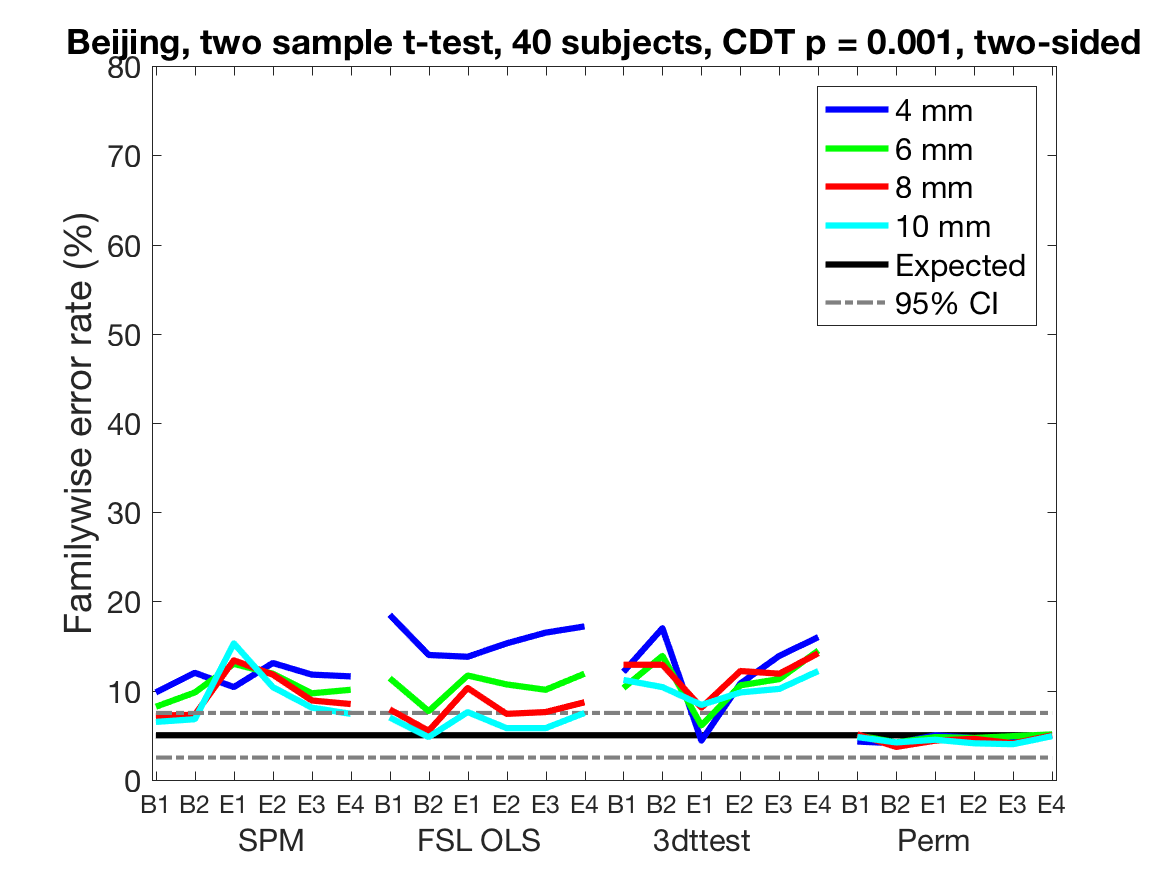}
}

\caption{\emph{A comparison of empirical familywise error rates for one-sided (left) and two-sided (right) tests, for a cluster defining threshold of p = 0.001. Designs B1 and B2 represent two block based activity paradigms, while E1, E2, E3 and E4 represent event related paradigms. Design E4 is randomized over subjects, while all other designs are the same for all subjects. The parametric methods perform worse for two one-sided tests at $\alpha_{\mathrm{FWE}}$ = 0.025, compared to a single one-sided test at $\alpha_{\mathrm{FWE}}$ = 0.05, while the permutation test produces nominal results in both cases. }}
\label{fig:fwe}
\end{figure*}

Figure~\ref{fig:fwe} shows estimated familywise error rates for one-sided and two-sided tests, where both should exhibit a nominal 5\% familywise false positive rate. The non-parametric permutation test produces similar results in both cases, while the parametric methods perform worse for two-sided tests.
 
\section{Discussion}

We have extended our original work on cluster false positive rates~\citep{eklundPNAS,eklundHBM} to two-sided tests, showing that parametric methods perform worse for two-sided tests. RFT p-values depend on a number of approximations:  
\begin{enumerate}
\item Joint normality over the image,
\item Sufficient smoothness for lattice images to behave like continuous processes,
\item Homogeneous smoothness (stationarity), so that the null distribution of cluster size does not vary over space,
\item Spatial dependence mostly local, i.e. the spatial autocorrelation function is proportional to a Gaussian density, and
\item Sufficiently high cluster-forming threshold so that the approximate distribution for cluster size is accurate.
\end{enumerate}
On this last assumption, the control of FWE depends on the accuracy of the cluster size distribution in its tail.  For example, it is of little consequence if the true cluster size FWE p-value is 0.6 and RFT estimates it as 0.5; in contrast, two-sided inference demands accuracy in the RFT approximation down to FWE 0.025, and then any inaccuracies are doubled as both positive and negative excursions are considered. In our findings, it appears that modest inaccuracies in the null cluster size distribution corresponding to FWE 0.05 (see Figure 1 (a), and general tendency to over estimate FWE) grow into larger inaccuracies when the more stringent FWE level 0.025 is used (the inference used twice for each result contributing to Figure 1 (b)).

In contrast, the non-parametric permutation test for a two-sample t-test is only based on the assumption of exchangeability between subjects, and therefore performs equally well for two one-sided tests at $\alpha_{\mathrm{FWE}}$ = 0.025.

\section*{Acknowledgements}

The authors have no conflict of interest to declare. This study was supported by Swedish research council grants 2013-5229 and 2017-04889. Funding was also provided by the Center for Industrial Information Technology (CENIIT) at Link\"{o}ping University, and the Knut and Alice Wallenberg foundation project "Seeing organ function". Thomas E. Nichols was supported by the Wellcome Trust (100309/Z/12/Z) and the NIH (R01 EB015611). The Nvidia Corporation, who donated the Nvidia Quadro P6000 graphics card used to run all permutation tests, is also acknowledged. This study would not be possible without the recent data-sharing initiatives in the neuroimaging field. We therefore thank the Neuroimaging Informatics Tools and Resources Clearinghouse and all of the researchers who have contributed with resting-state data to the 1,000 Functional Connectomes Project.

\bibliography{references_full}

\begin{thebibliography}{7}
\expandafter\ifx\csname natexlab\endcsname\relax\def\natexlab#1{#1}\fi
\providecommand{\bibinfo}[2]{#2}
\ifx\xfnm\relax \def\xfnm[#1]{\unskip,\space#1}\fi
\bibitem[{Biswal et~al.(2010)Biswal, Mennes, ... \& Milham}]{biswal2}
\bibinfo{author}{Biswal, B.}, \bibinfo{author}{Mennes, M.},
  \bibinfo{author}{..., X.~Z.}, \& \bibinfo{author}{Milham, M.}
  (\bibinfo{year}{2010}).
\newblock \bibinfo{title}{Toward discovery science of human brain function}.
\newblock {\it \bibinfo{journal}{PNAS}\/},  {\it \bibinfo{volume}{107}\/},
  \bibinfo{pages}{4734--4739}.
\bibitem[{Chen et~al.(2018)Chen, Cox, Glen, Rajendra, Reynolds \&
  Taylor}]{twosided}
\bibinfo{author}{Chen, G.}, \bibinfo{author}{Cox, R.~W.},
  \bibinfo{author}{Glen, D.~R.}, \bibinfo{author}{Rajendra, J.~K.},
  \bibinfo{author}{Reynolds, R.~C.}, \& \bibinfo{author}{Taylor, P.~A.}
  (\bibinfo{year}{2018}).
\newblock \bibinfo{title}{{A tail of two sides: Artificially doubled false
  positive rates in neuroimaging due to the sidedness choice with t-tests}}.
\newblock {\it \bibinfo{journal}{Human Brain Mapping}\/}, .
\bibitem[{Cox et~al.(2017)Cox, Chen, Glen, Reynolds \& Taylor}]{coxbiorxiv}
\bibinfo{author}{Cox, R.}, \bibinfo{author}{Chen, G.}, \bibinfo{author}{Glen,
  D.}, \bibinfo{author}{Reynolds, R.}, \& \bibinfo{author}{Taylor, P.}
  (\bibinfo{year}{2017}).
\newblock \bibinfo{title}{{FMRI Clustering in AFNI: False-Positive Rates
  Redux}}.
\newblock {\it \bibinfo{journal}{Brain Connectivity}\/},  {\it
  \bibinfo{volume}{7}\/}, \bibinfo{pages}{152--171}.
\bibitem[{Cox(2018)}]{etac}
\bibinfo{author}{Cox, R.~W.} (\bibinfo{year}{2018}).
\newblock \bibinfo{title}{Equitable thresholding and clustering}.
\newblock {\it \bibinfo{journal}{bioRxiv, 10.1101/295931}\/}, .
\bibitem[{Eklund et~al.(2018)Eklund, Knutsson \& Nichols}]{eklundHBM}
\bibinfo{author}{Eklund, A.}, \bibinfo{author}{Knutsson, H.}, \&
  \bibinfo{author}{Nichols, T.} (\bibinfo{year}{2018}).
\newblock \bibinfo{title}{Cluster failure revisited: impact of first level
  design and physiological noise on cluster false positive rates}.
\newblock {\it \bibinfo{journal}{Human Brain Mapping}\/}, .
\bibitem[{Eklund et~al.(2016)Eklund, Nichols \& Knutsson}]{eklundPNAS}
\bibinfo{author}{Eklund, A.}, \bibinfo{author}{Nichols, T.}, \&
  \bibinfo{author}{Knutsson, H.} (\bibinfo{year}{2016}).
\newblock \bibinfo{title}{Cluster failure: why {fMRI} inferences for spatial
  extent have inflated false positive rates}.
\newblock {\it \bibinfo{journal}{PNAS}\/},  {\it \bibinfo{volume}{113}\/},
  \bibinfo{pages}{7900--7905}.
\bibitem[{Winkler et~al.(2014)Winkler, Ridgway, Webster, Smith \&
  Nichols}]{winkler}
\bibinfo{author}{Winkler, A.}, \bibinfo{author}{Ridgway, G.},
  \bibinfo{author}{Webster, M.}, \bibinfo{author}{Smith, S.}, \&
  \bibinfo{author}{Nichols, T.} (\bibinfo{year}{2014}).
\newblock \bibinfo{title}{Permutation inference for the general linear model}.
\newblock {\it \bibinfo{journal}{NeuroImage}\/},  {\it \bibinfo{volume}{92}\/},
  \bibinfo{pages}{381--397}.

\end{thebibliography}

\end{document}